\documentstyle[prd,aps,epsfig]{revtex}

\def\npb#1#2#3{    {\it Nucl. Phys. }{\bf B #1} (19#2) #3}
\def\plb#1#2#3{    {\it Phys. Lett. }{\bf B #1} (19#2) #3}
\def\prd#1#2#3{    {\it Phys. Rev. }{\bf D #1} (19#2) #3}

\def\prl#1#2#3{    {\it Phys. Rev. Lett. }{\bf #1} (19#2) #3}
\def\ptp#1#2#3{    {\it Prog. Theor. Phys. }{\bf #1} (19#2) #3}
\def\rmp#1#2#3{    {\it Rev. Mod. Phys. }{\bf #1} (19#2) #3}
\def\zpc#1#2#3{    {\it Z. Physik }{\bf C #1} (19#2) #3}

\def\ibid#1#2#3{   {\it ibid. }{\bf #1} (19#2) #3}

\begin{document}
\draft
\twocolumn[\hsize\textwidth\columnwidth\hsize\csname
@twocolumnfalse\endcsname

\title{Do experiments suggest a hierarchy problem?}
\author{Francesco Vissani}
\address{International Centre for Theoretical Physics,\\
Strada Costiera 11, I-34013 Trieste, Italy}
\date{September 18, 1997}
\maketitle
\begin{abstract}
The hierarchy problem of 
the scalar sector of the 
standard model is reformulated,
emphasizing the role of experimental 
facts that may suggest the existence 
of a new physics large mass scale,
for instance indications of the 
instability of the matter,
or indications in favor of massive neutrinos.
In the see-saw model for the neutrino masses
a hierarchy problem arises if the mass of 
the right-handed neutrinos 
is larger than approximatively $10^7$ GeV: 
this problem, and its possible solutions, 
are discussed.
\end{abstract}
\pacs{PACS: 12.60.-i; 14.60.St; 11.30.Pb}
\vskip0.4truecm] 
\noindent 1. We speak of a hierarchy problem when
two largely different energy scales are present in the theory, 
but there is no symmetry that stabilizes 
the light scale from corrections coming from the large scale \cite{gh}.

This problem is commonly invoked to argue 
against the simple structure of the Higgs potential of the
standard model, since the massive parameter $\mu^2$ 
appearing as $-\mu^2 |H|^2$ 
in the potential (the light scale) 
can in principle receive corrections from any larger scale. 
Which kind of mass the problem pertains? 
It can be formulated in terms 
of the renormalized mass, let us say in $\overline{MS}$ 
scheme, noticing that at external momenta  
above a heavy threshold scale 
$M_{heavy}$ the parameter $\mu^2$
will acquire loop contributions 
of the order of $M_{heavy}^2,$ 
times the coupling to the heavy particle.
In this case, the renormalization group flow 
in the standard model is
unnatural, in the sense that 
the initial conditions at some large scale have to
be extremely fine-tuned to reproduce a Higgs mass
below the TeV scale, if the coupling of the Higgs particle with
the particle of mass $M_{heavy}$ is not very small.
{}From another point of view, it was remarked that the 
bare scalar mass receives quadratic corrections,
if the theory is regulated with a cutoff in the momenta
\cite{quadraticDivergences}.
This may be considered a less relevant aspect, since 
the standard model is a renormalizable theory, and
there is no way to give sense to bare 
parameters in this context;
the cutoff can be thought of as a technical device, and
in last analysis, other regulators can be chosen.

Notice that to speak of 
a ``problem'' one is taking a 
theoretical
point of view: One does not like to assume, without
motivations, that a hypothetical
fundamental theory that should explain 
the observed quantities
and the various
parameters of the standard model 
should be forced to 
have a fine-tuning like 
the one discussed above.
This principle can be used to 
select possible extensions 
of the standard model:
What is needed is simply stating 
a quantitative criterion of naturalness 
(such a program was formulated in \cite{nat}).

Once this principle is accepted,
the discussion about 
its actual relevance 
is reduced to 
two experimental terms.
The first is if 
a fundamental Higgs particle exists. 
Assuming that it exists, 
we face the other aspect:
before speaking of a hierarchy problem,
one has to understand
if there are signals of physics beyond
the standard model, that, in turn, 
point to the existence of larger energy scales.

We will not rely on the Planck mass scale 
in the following discussion, 
since in our opinion the formulation of
a quantum theory of the gravity is in a 
preliminary stage, 
and the experimental perspectives
are unclear \cite{gm}.
We want instead to discuss the
relevance of signals of violations of the 
global symmetries of 
the standard model, the baryon
and the lepton numbers B and L, paying attention to 
the experimental perspectives 
that we can foresee at present. 
\vskip0.4truecm\noindent 2. Let us start to discuss 
possible signals of matter instability.
If discovered, they 
would strongly suggest the 
existence of a large mass scale,
most probably related to a deeper  
layer of gauge unification
(the alternative hypothesis 
of light mediators of matter instability, very weakly
coupled with the matter, should be seriously considered 
if B+L conserving nucleon decay modes would be observed 
\cite{wwz,d7}).
Suppose that proton 
decay signals would be within reach,
say at Superkamiokande.
To be concrete, 
let us imagine the case in which 
the decay channels involving strange mesons are
the dominating ones, that may
indicate us that the
physics responsible of 
the proton decay and of the 
origin of the (family hierarchical) 
fermion masses is the same.
Assuming that the couplings involved
in the decay are of the order of a typical Yukawa 
couplings $m_s/v\approx 10^{-3},$ 
a sufficient suppression of 
the nucleon lifetime can be obtained only if the 
mass of the mediator $M_X$ is 
close to $10^{12}$ GeV  
(we assumed: $\Gamma_p\sim M^{-4}_X.$)  
Therefore, $\mu^2$ receives the contribution 
$\delta\mu^2\approx y^2 M_X^2/(4 \pi)^2$ that 
is much larger than 1 TeV$^2,$ unless
the effective coupling $y$
of the light Higgs with the heavy particle
is very small, approximatively $y<10^{-8}.$ 
It is easy to understand that 
for a typical theoretical scheme
(in which $y$ can appear at 
one-loop or even at tree level)  
the contributions to $\delta\mu^2$ 
can be very large.  
In conclusion, this
scenario would probably require 
us to wonder about the hierarchy problem and 
about its solution.

It is remarkable that the 
supersymmetric extensions of the
standard model, with masses of the 
supersymmetric particles around
the electroweak scale,
are able to offer 
a way out from the 
hierarchy problem due to the 
non-renormalization theorem 
\cite{nonren} 
and at the same time
are compatible with the hypothesis of a minimal SU(5) 
unification group structure at an energy scale around 
$2\times 10^{16}$ GeV \cite{dg,gcc}.
This may be regarded as {\em the} 
solution \cite{supersymmetry}, but in the 
present stage of development
it is not clear if a gauge hierarchy 
problem has to be addressed, 
since no signal of matter 
instability has been found yet.
In this connection, it is important to remark that
supersymmetric grand unified models that 
predict that nucleon decay signal may be
within reach (in the close future) 
have been indeed proposed \cite{protdec}.
However, one should not forget that some 
supersymmetric grand unified model can be 
already excluded by
present experimental information 
on matter stability\cite{pd}, 
or, on the opposite extreme, that some model
entails an essentially stable nucleon \cite{gia}.
Even if somewhat disappointing,
it may be fair to say that 
this is due to the fact that the  
``supersymmetric grand 
unification'' is still 
not a completely defined program. 
Coming back to the 
main focus of the 
present work,
we conclude that (despite the 
theoretical promises) the 
experimental studies of
matter stability
do not permit us at present 
to infer the existence
of a hierarchy problem.

\vskip0.4truecm\noindent 3. There is, however, 
an independent way of arguing of
a hierarchy problem 
in certain extensions of the standard model. 
This argumentation is based on 
the presence of non-zero neutrino masses,
that could imply the solution of 
long standing problems 
with solar neutrino flux,
and may be confirmed by the next 
round of experiments. 

It is in principle possible
that also the neutrino 
masses are related
to a new gauge structure 
manifesting itself at higher scales;
if this would be true, 
we should again face 
a gauge hierarchy problem 
\cite{gaugeNeutrino}.
However, we want to be conservative 
in the assumptions. 
So, instead of jumping to conclusions, 
we address the question: What can we learn, 
using the indications of non-zero neutrino masses,
on the structure of the theory that should 
extend the standard model?

Let us consider the
see-saw model for neutrino masses \cite{seesaw}.
The heavy right-handed neutrinos, with mass $M_R,$ couple
with the Yukawa coupling $y_\nu$ to the 
left-handed neutrinos, and give
them a mass $m_\nu = (y_\nu\, v)^2/M_R$
($v=174$ GeV).
In non supersymmetric theories the renormalized
mass $\mu^2$ will receive corrections order 
\begin{equation}
\delta \mu^2\approx 
\frac{y_\nu^2}{(2\pi)^2} \ M_R^2\ \log(q/M_R) ,
\label{corrections}
\end{equation}
for momenta $q$ larger 
than $M_R$ (see fig.\ 1).
We can rewrite these corrections as:
\begin{equation}
\delta \mu^2\approx 
\frac{m_\nu M_R^3}{(2\pi\, v)^2} \log(q/M_R) .
\label{fine}
\end{equation}
Equation (\ref{fine}) points to the
hierarchy problem that is inherent to 
the see-saw models for the neutrino 
masses.
\begin{figure}
\centerline{\epsfig{file=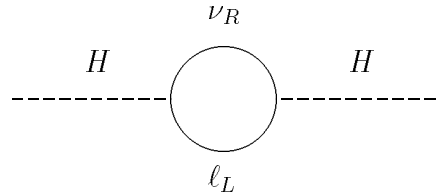}}
\caption{The Feynman diagram originating the corrections 
in eq.\ (\ref{corrections}); $\nu_R$ denotes
the right-handed neutrino of mass $M_R,$
$\ell_L=(\nu_L,e_L)$ the leptonic- and $H$ the Higgs-doublets.}
\end{figure}

We specify the previous formula in two concrete cases,
considering neutrinos that may be relevant to the
solution of the 
solar neutrino problem and may serve as 
the hot dark matter (HDM)
candidate respectively.   
Assuming small mixing, the contribution to $\mu^2$ 
will not exceed 1 TeV$^2$ if the following upper bounds hold true:
\begin{eqnarray}
{m_\nu({\rm solar}) = 3\times 10^{-3}\ {\rm eV}} & 
\Longrightarrow 
& M_R {\ \raisebox{-.4ex}{\rlap{$\sim$}} 
\raisebox{.4ex}{$<$}\ } 7.4\times 10^7\ {\rm GeV}\nonumber \\
{m_\nu({\sc hdm}) = 6\ {\rm eV}}\ \ \ \ \ \ \ \ \ \, & 
\Longrightarrow 
& M_R {\ \raisebox{-.4ex}{\rlap{$\sim$}} 
\raisebox{.4ex}{$<$}\ } 5.8\times 10^6\ {\rm GeV} .
\label{two}
\end{eqnarray}
In the previous estimation we assumed the logarithm 
of order unity (in other terms, 
we are using the criterion of naturalness: 
$d\mu^2/d\log q{\ \raisebox{-.4ex}{\rlap{$\sim$}} 
\raisebox{.4ex}{$<$}\ } 1 {\rm TeV}^2$).
Let us stress that the figures in eq.\ (\ref{two}) 
should be taken as indicative, 
since we assumed  that the 
mixing angles and the phases
in the lepton matrices are small;
their presence can modify to a certain extent 
the relation between the masses of 
the light and the heavy neutrinos.
However, for given values of the left-handed and right-handed
neutrino masses, the radiative contribution 
to $\mu^2$ tend to increase in presence of mixing and phases.

Under the same assumptions, 
the conditions (\ref{two}) on $M_R$
are equivalent to upper bounds
on the Yukawa 
couplings:
\begin{eqnarray}
m_\nu({\rm solar}) = 3\times 10^{-3}\ {\rm eV} & 
\Longrightarrow 
& y_\nu {\ \raisebox{-.4ex}{\rlap{$\sim$}} 
\raisebox{.4ex}{$<$}\ } 8.5\times 10^{-5} \nonumber \\
m_\nu({\sc hdm}) = 6\ {\rm eV} \ \ \ \ \ \ \ \ \ \, & 
\Longrightarrow 
& y_\nu {\ \raisebox{-.4ex}{\rlap{$\sim$}} 
\raisebox{.4ex}{$<$}\ } 1.1\times 10^{-3} .
\label{twotwo}
\end{eqnarray}
For comparison, 
we notice that if $M_R\approx 1$ TeV
(of interest for search at accelerators)
the Yukawa couplings are
$y_\nu\approx  3.1\times 10^{-7}, 1.4\times 10^{-5}$ 
in the two cases considered.

Therefore, to be able to assess the presence 
of a hierarchy problem, we still lack the information 
on the scale of the Majorana neutrinos $M_R,$ 
or on the size of the Yukawa couplings.
A recent discussion \cite{Smirnov} of the structure of the
right-handed mass matrix in the see-saw model suggests
masses larger than those in (\ref{two}).
Notice however that the underlying assumption is
the unification  of the Yukawa couplings
of the neutrinos and of the 
up-type-quarks; for smaller neutrino Yukawa coupling, 
lighter $M_R$ are needed.
For instance, this is what 
happens if neutrinos are Dirac particles,
that is when $M_R \ll m_\nu$ (and there is no
direct Majorana mass term); 
the neutrino mass reduces to $y_\nu\, v ,$ 
and the Yukawa couplings are
very small ($y_\nu =  1.7\times 10^{-14}$
for solar neutrinos and
$y_\nu = 3.4\times 10^{-11}$
for HDM component neutrinos).

An interesting information on the Yukawa couplings  
follows if we assume the Fukugita-Yanagida scenario 
for baryogenesis \cite{FY1} (see also \cite{FY2,FY3,FY4}), 
in which the decay of the lightest right-handed neutrino,
of mass $M_{R\,l},$ originates a lepton asymmetry
that, in a second stage, can be converted in the 
presently observed baryon asymmetry.
In fact, this scenario can be realized 
if the Yukawa couplings
provide sufficient mixing with 
a heavier neutrino of mass $M_{R\,h}:$
\begin{equation}
\frac{M_{R\,l}}{M_{R\,h}} \ 
\frac{{\rm Im}[({Y_\nu}^\dagger 
Y_\nu)_{hl}^2]}{({Y_\nu}^\dagger Y_\nu)_{ll}} 
\approx 10^{-5} ,
\label{mix}
\end{equation}
in the case of {\em hierarchical} 
masses of right-handed neutrinos,
as discussed in \cite{FY3}.
Considering the inequality:
$|({Y_\nu}^\dagger Y_\nu)_{hl}|^2 \le 
({Y_\nu}^\dagger Y_\nu)_{hh}\ 
({Y_\nu}^\dagger Y_\nu)_{ll},$
that follows from the non-negativity of the matrix 
${Y_\nu}^\dagger Y_\nu,$ we obtain: 
\begin{equation}
10^{-5} {\ \raisebox{-.4ex}{\rlap{$\sim$}} 
\raisebox{.4ex}{$<$}\ } ({Y_\nu}^\dagger Y_\nu)_{hh} .
\label{constr}
\end{equation} 
Comparing with  eq.\ (\ref{twotwo}),
we come to the conclusion that
the corrections to $\mu^2$ exceed the TeV$^2;$
in other terms, eq.\ (\ref{constr}) 
suggests the vicinity of 
a hierarchy problem.

This conclusion is related to a conjectural mechanism
for baryogenesis, that however is quite natural once the
existence of right-handed neutrinos has been assumed.
For this reason, it is of interest 
to search for a loophole in the above argument.
Let us therefore abandon the hypothesis of hierarchical
right-handed neutrinos, 
and contemplate the case in which these particles  
are nearly degenerate;
it turns out that 
the estimation (\ref{mix}) 
is no longer correct.
In fact, the lepton asymmetry 
produced in the decay 
is dominated by the ``wavefunction'' 
contribution \cite{FY3,FY4}, that increases for  
smaller mass splitting, and that eventually reaches
its maximum when the splitting 
is comparable to the decay widths
of the right-handed neutrinos \cite{FY4}.
This makes it possible to reproduce 
the observed baryon number 
with smaller Yukawa couplings 
than those implied by eq.\ (\ref{constr}), and
gives a chance to avoid the hierarchy problem in
the minimal framework we are considering.
We will not address the question of the 
theoretical likelihood of this
very constrained scenario for neutrino masses.
However, it is important
to stress again that even in this framework  
the right-handed neutrinos 
would be 
relatively light (eq.\ (\ref{two})).

\vskip0.4truecm\noindent 4. Finally, we discuss 
possible solutions of the
hierarchy problem that arises if 
the see-saw model is the true theory of
the neutrino masses, and the 
right-handed masses are large in comparison 
with eq.\ (\ref{two}) (as suggested 
by eq.\ (\ref{constr}), modulo the {\em caveats} above). 
In this case, one could
advocate for supersymmetry 
at low energy on the basis of 
the criterion of naturalness.
We recall the argument:
The quadratic corrections
to the massive parameters of the Higgs potential
entail in supersymmetric theories 
$M_R^2-\widetilde{M}_R^2 , $ 
the mass splitting of the 
right-handed neutrinos and their 
scalar partners instead of $M_R^2$
(compare with eq.\ (\ref{corrections}));  
the natural expectation is that 
$M_R^2-\widetilde{M}_R^2{\ \raisebox{-.4ex}{\rlap{$\sim$}} 
\raisebox{.4ex}{$<$}\ } 1{\rm TeV}^2,$ due, 
for instance, to a relation of 
this mass splitting and the splitting between
the charged leptons and their scalar 
counterparts.
As a conclusion, the presence of the large
mass scale $M_R^2\gg 1{\rm TeV}^2$ does not 
imply any hierarchy problem.

In this supersymmetric context, we remark that 
the mass splitting  $M_R^2-\widetilde{M}_R^2 $ 
could affect via one-loop corrections
the value of the lightest Higgs mass,
in close similarity with what happens due to 
the top-stop corrections \cite{topstop}.
In fact, these loop corrections are  
of the same nature of the corrections to $\mu^2$ 
discussed in eq.\ (\ref{corrections}).

Of course, the 
argument for supersymmetry
is far-reaching,
and does not apply only to the 
see-saw model. In fact, once the low energy 
supersymmetry hypothesis is accepted,
the light scales are ``protected'' 
against the presence of the heavy scales,  
and the theoretical speculations 
involving very high energy scales
do not meet these types of problem.
The urgency of the remarks above   
stays in the consideration that
the strongest indications in favor of physics 
beyond the standard model come from neutrino physics.

If the model of the neutrino masses
is not the see-saw model we have 
other possibilities to elude 
the hierarchy problem:
We can assume that the scale 
at which the neutrino masses are generated 
is not far from the electroweak one.
This can happen in the 
models in which the smallness of the 
neutrino masses is related to 
loop effects \cite{Petcov}.
Even in the 
context of minimal supersymmetric 
models (in particular without right-handed neutrinos) 
other mechanisms for the generation 
of the neutrino masses are possible.
We are referring to the R-parity breaking models, in
which {\em a priori} large violations 
of the lepton number may 
be present \cite{Rspont,Rexpl}.
Again, the crucial remark is that 
in these models no large scale 
(besides the scale of 
the supersymmetric particles) is present.
Can we distinguish this possibility?
If the neutrino masses originate in these kinds
of models, the expectation is that
other signals of R-parity breaking should show up \cite{rm}. 

\vskip0.4truecm\noindent 5. To summarize, 
massive neutrinos 
point to a hierarchy problem in 
possible extensions of the 
standard model, independently from 
the assumption of grand 
unification.
We discussed how
this remark may result in 
an argument in favor
of certain theoretical models.

\acknowledgments{It is a pleasure to 
acknowledge discussions with 
K.S.\ Babu,
H.\ Minakata,
E.\ Roulet,
G.\ Senjanovi\'c
and
A.Yu.\ Smirnov.}


\begin{thebibliography}{99}
\bibitem{gh} 
This problem has been stated in connection 
with gauge hierarchies by\\
E. Gildener, \prd{14}{76}{1667}.
\bibitem{gm} However, L. Gonzalez-Mestres in {\tt physics/9705031}
suggested 
the possible relevence of Planck scale physics  
for the cosmic rays of highest energy.
\end{thebibliography}
\end{document}